\documentclass[useAMS,usenatbib]{mn2e}
\usepackage{graphicx}
\usepackage{color}
\usepackage{ulem}

\newcommand{\nbf}{}
\newcommand{\DM}{D_M}
\newcommand{\DV}{D_V}

\newcommand{\DL}{d_L}
\newcommand{\sig}{\sigma}
\newcommand{\al}{\alpha}
\newcommand{\Om}{\Omega_m}

\newcommand{\zprime}{z^\prime}
\newcommand{\LCDM}{$\Lambda\mbox{CDM}$}

\title[Consistency and constraints with intermediate-redshift datasets]
{First-year Sloan Digital Sky Survey-II (SDSS-II) supernova results: 
consistency and constraints with other intermediate-redshift datasets}

\author[H. Lampeitl et al.]{H.~Lampeitl$^{1}$\thanks{E-mail: Hubert.Lampeitl@port.ac.uk}, 
R.C.~Nichol$^{1}$,
H.-J.~Seo$^{2}$, 
T.~Giannantonio$^{1,3}$,
C.~Shapiro$^{1}$,\newauthor
B.~Bassett$^{4,5}$,
W.J.~Percival$^{1}$,
T.M.~Davis$^{6,7}$,
B.~Dilday$^{8}$,
J.~Frieman$^{2,9}$, \newauthor
P.~Garnavich$^{10}$,
M.~Sako$^{11}$,
M.~Smith$^{1,4}$,
J.~Sollerman$^{7,12}$,
A.C.~Becker$^{13}$, \newauthor
D.~Cinabro$^{14}$,
A.V.~Filippenko$^{15}$,
R.J.~Foley$^{15,16,17}$,
C.J.~Hogan$^{2}$, 
J.A.~Holtzman$^{18}$, \newauthor 
S.W.~Jha$^{8}$,
K.~Konishi$^{19,20}$, 
J.~Marriner$^{2}$, 
M.W.~Richmond$^{21}$,
A.G.~Riess$^{22,23}$, \newauthor
D.P.~Schneider$^{24}$, 
M.~Stritzinger$^{7,25}$, 
K.J.~van der Heyden$^{26}$,
J.T.~VanderPlas$^{13}$, \newauthor
J.C. Wheeler$^{27}$, 
C.~Zheng$^{28}$\\
\vspace{1cm}
\\
$^{1}$Institute of Cosmology and Gravitation, University of Portsmouth, Portsmouth, PO1 3FX, UK\\
$^{2}$Center for Particle Astrophysics, Fermi National Accelerator Laboratory, P.O. Box 500, Batavia, IL 60510, USA\\
$^{3}$Argelander-Institut f\"{u}r Astronomie, Universit\"{a}t Bonn, Auf dem H\"{u}gel 71, D-53121 Bonn, Germany\\
$^{4}$Department of Mathematics and Applied Mathematics, University of Cape Town, Rondebosch 7701, South Africa\\
$^{5}$South African Astronomical Observatory, P.O. Box 9, Observatory 7935, South Africa\\
$^{6}$ School of Mathematics and Physics, University of Queensland, Brisbane QLD 4072, Australia\\
$^{7}$Dark Cosmology Centre, Niels Bohr Institute, University of Copenhagen, DK-2100, Copenhagen, Denmark\\
$^{8}$Department of Physics and Astronomy, Rutgers the State University of New Jersey, 136 Frelinghuysen Road, Piscataway, NJ 08854, USA\\
$^{9}$Kavli Institute for Cosmological Physics, The University of Chicago, 5640 South Ellis Avenue Chicago,IL 60637, USA\\
$^{10}$Physics Department, University of Notre Dame, Notre Dame, IN 46556, USA\\
$^{11}$University of Pennsylvania, Department of Physics and Astronomy, 209 South 33rd Street, Philadelphia, PA 19096, USA\\
$^{12}$Department of Astronomy, AlbaNova, Stockholm University, SE-106 91 Stockholm, Sweden\\
$^{13}$Department of Astronomy, University of Washington, Box 351580, Seattle, WA 98195, USA\\
$^{14}$Wayne State University, Department of Physics and Astronomy, Detroit, MI 48202, USA\\
$^{15}$Department of Astronomy, University of California, Berkeley, CA 94720-3411, USA\\
$^{16}$Harvard-Smithsonian Center for Astrophysics, 60 Garden Street, Cambridge, MA 02138\\
$^{17}$Clay Fellow\\
$^{18}$Dept. of Astronomy, MSC 4500, New Mexico State University, P.O. Box 30001, Las Cruces, NM 88003, USA\\
$^{19}$Institute for Cosmic Ray Research, The University of Tokyo, 5-1-5 Kashiwa, Kashiwa City, Chiba 277-8582, Japan\\
$^{20}$Department of Physics, Graduate School of Science, The University of Tokyo, 7-3-1 Hongo, Bunkyo, Tokyo 113-0033, Japan\\
$^{21}$Physics Department, Rochester Institute of Technology, Rochester, NY 14623, USA\\
$^{22}$Department of Physics and Astronomy, Johns Hopkins University, Baltimore, MD 21218, USA\\
$^{23}$Space Telescope Science Institute, 3700 San Martin Drive, Baltimore, MD 21218, USA\\
$^{24}$Department of Astronomy and Astrophysics, 525 Davey Laboratory, Pennsylvania State University, University Park, PA 16802, USA\\
$^{25}$Las Campanas Observatory, Carnegie Observatories, Casilla 601, La Serena, Chile\\ 
$^{26}$Astronomy Department, University of Cape Town, Private Bag X3, Rondebosch 7701, South Africa\\
$^{27}$Department of Astronomy, McDonald Observatory, University of Texas, Austin, TX 78712, USA\\
$^{28}$Kavli Institute for Particle Astrophysics and Cosmology, Stanford University, CA 94305-4060
}
\begin{document}

\date{Accepted --------. Received --------------}

\pagerange{\pageref{firstpage}--\pageref{lastpage}} \pubyear{2002}

\maketitle

\label{firstpage}

\newpage
\clearpage
\begin{abstract}
We present an analysis of the luminosity distances of Type Ia Supernovae from 
the Sloan Digital Sky Survey-II (SDSS-II) Supernova Survey in conjunction with other intermediate redshift ($z<0.4$) cosmological 
measurements including redshift-space distortions from the Two-degree Field Galaxy Redshift Survey (2dFGRS), the Integrated 
Sachs-Wolfe (ISW) effect seen by the SDSS, and the latest Baryon Acoustic Oscillation (BAO) distance scale from both the SDSS 
and 2dFGRS. We have analysed the SDSS-II SN data alone using a variety of ``model-independent" methods and find evidence 
for an accelerating universe at  $>$97\% level from this single dataset. We find good agreement between the supernova and BAO 
distance measurements, both consistent with a $\Lambda$--dominated CDM cosmology, as demonstrated through an analysis  of 
the distance duality relationship between the luminosity ($d_L$) and angular diameter ($d_A$) distance measures. 
We then use these data to estimate $w$ within this restricted redshift range ($z<0.4$). Our most stringent result comes from the 
combination of all our intermediate--redshift data (SDSS-II SNe, BAO, ISW and redshift--space distortions), giving 
$w = -0.81^{+0.16}_{-0.18}(\mathrm{stat}){\pm0.15}(\mathrm{sys})$ and $\Omega_M=0.22^{+0.09}_{-0.08}$ assuming a flat universe. 
This value of $w$, and associated errors, only change slightly if curvature is allowed to vary, consistent with constraints from the 
Cosmic Microwave Background. We also consider more limited combinations of the geometrical (SN, BAO) and dynamical 
(ISW, redshift-space distortions) probes.
\end{abstract}

\begin{keywords}
(stars:) supernovae: general, (cosmology:) observations, distance-scale, cosmological parameters, large-scale structure of universe
\end{keywords}

\section{Introduction}

It is now widely believed that the late-time expansion of the universe is accelerating.  General Relativity (GR) implies that the acceleration 
is driven by ``dark energy" (DE) --- an unknown energy component in the universe with a negative effective pressure.  Describing dark energy by 
an equation-of-state parameter of $w(z) = p/(\rho c^2 )$ requires that $w<-1/3$.  Alternatively, accelerated expansion could be an indication that GR is not the 
correct theory of gravity or that we have applied GR incorrectly in a cosmological context (see recent reviews of DE by \citealt{2003RvMP...75..559P}, 
\citealt{2006astro.ph..5313U}, \citealt{2007AIPC..957...21C}, and \citealt{2008arXiv0803.0982F}).

Over the last decade, the most direct way of studying this acceleration of the expansion of the universe, and therefore DE, has been using Type Ia Supernovae (SNe), as they have been shown by many authors to be well-calibrated ``standard candles" in the universe, i.e., their relative distances can be 
determined from the dependence of their peak luminosity on
the shape of the light curve. This method was used to great effect by astronomers in 1998 to provide the first evidence for an accelerated universe (\citealt{1998AJ....116.1009R, 1999ApJ...517..565P}; see \citealt{2005ASSL..332...97F} for a review).

Briefly, a Type Ia supernova occurs when a white dwarf star in a close binary system accretes enough mass from its companion to undergo a thermonuclear explosion in the core. Both \citet{1993ApJ...413L.105P} and
\citet{1993PASP..105..787H} have shown that such explosions can serve as consistent
light sources in the universe to high accuracy. This is achieved  by transforming the measured light curve of the explosion into the rest frame of the supernova (so called K-corrections) and then correcting the luminosity at maximum as a function of the shape of the rest-frame light curve.

Several techniques now exist for fitting SN light curves known under different acronyms 
($\Delta m_{15}$, \citealt{1996AJ....112.2391H};
MLCS, \citealt{1996ApJ...473...88R};
stretch, \citealt{1997ApJ...483..565P};
CMAGIC, \citealt{2003ApJ...590..944W};
BATM, \citealt{2003ApJ...594....1T}; 
SALT, \citealt{2005A&A...443..781G}; 
$\Delta C_{12}$, \citealt{2006ApJ...645..488W};
SALT2, \citealt{2007A&A...466...11G};
SiFTO, \citealt{2008ApJ...681..482C}). 
In this analysis, we consider MLCS2k2 \citep{2007ApJ...659..122J, 1996ApJ...473...88R}, which is among the most
commonly used, and best tested, of these various techniques. 

Recently, several dedicated SN surveys have been carried out to confirm and extend the earlier detections of an accelerating universe (HST, 
\citealt{2004ApJ...607..665R,2007ApJ...659...98R};
SNLS, \citealt{2006A&A...447...31A}; ESSENCE, \citealt{2007ApJ...666..694W}) as well as new compilations of existing SN datasets 
\citep{2007ApJ...666..716D, 2008arXiv0804.4142K, 2009arXiv0901.4804H}. In addition to supernovae, observations of Baryon Acoustic Oscillations (BAO) can be used to measure distances in the universe \citep{2003ApJ...594..665B, 2003ApJ...598..720S, 2003PhRvD..68f3004H}. The BAO are caused by sound waves in the early universe which leave a preferred scale in the distribution of matter equal to the sound horizon at recombination. Today, this scale corresponds to $\sim100/h$ Mpc (Hubble constant at present: $H_0 = 100h$~km/s/Mpc) and can thus be used as a ``standard ruler" throughout the universe. 
The BAO signature has been detected in the clustering of galaxy clusters by \citet{2001ApJ...555...68M}, in the correlation of galaxies in the 
Sloan Digital Sky Survey (SDSS, \citealt{2000AJ....120.1579Y}) by 
\citet{2005ApJ...633..560E}, \citet{2006AA...459..375H}, \citet{2007MNRAS.378..852P}, and \citet{2007MNRAS.374.1527B},
and in the Two-degree Field Galaxy Redshift Survey (2dFGRS, \citealt{2001MNRAS.328.1039C}) by \citet{2005MNRAS.362..505C}.

In addition to the geometrical methods discussed above, observations of the dynamical properties of matter can provide constraints on the matter density of the universe and $w$ (assuming General Relativity is the appropriate theory of gravity). For example, the growth rate of structure in the universe can be observed via the coherent infall of galaxies into large clusters and superclusters of galaxies seen in redshift surveys \citep{1987MNRAS.227....1K}. Also, the growth of structure can be measured using the late-time Integrated Sachs-Wolfe (ISW) effect \citep{1967ApJ...147...73S}, which has now been detected to high significance through the cross-correlation of galaxy surveys with the Cosmic Microwave Background (CMB) (see \citealt{2008PhRvD..77l3520G} and references therein). 
The ISW is sensitive to deviations from a matter-dominated, Einstein-de Sitter universe ($\Omega_M=1$, where $\Omega_M$ is the matter density
at present divided by the critical density.).

Taken together, the present combination of cosmological measurements suggests we live in a flat univers, dominated by a cosmological constant ($\Lambda$) with the energy density in matter and $\Lambda$ known to a statistical accuracy of better than a few percent \citep[see][]{2009ApJS..180..306D}. However, several of these cosmological measurements, especially SNe, are becoming limited by their systematic uncertainties which are now dominating, e.g., \citet{2009arXiv0901.4804H} showed that the best combination of available SNe and BAO measurements provide $1+w=0.013^{+0.066}_{-0.068}$ but with a systematic uncertainty of $0.11$. 
Therefore, it is clear that future cosmological surveys must resolve these systematic errors through new observations and better analysis methods to mitigate their effect.  

This paper is one of three complementary papers focused on the cosmological analysis of a new sample of  intermediate supernova distances recently obtained by the SDSS-II Supernova Survey (see Section \ref{chapter_1st_year_data} for details). Our analysis differs from those presented in our companion papers of \citet{kessler} and \citet{davis2008}, as we first study the cosmological information obtained solely from the SDSS-II SN sample, and then in combination with other cosmological probes over the same redshift range ($z<0.4$).  Alternatively,  \citet{kessler} presents a detailed examination of the impact of both statistical and systematic errors on deriving standard cosmological constraints based on the combination of the SDSS-II SN with most of the currently available high and low redshift SNe and which are all analysed in a consistent way. \citet{davis2008} then use the same compilation of data to study an expanded set of exotic cosmological models, in combination with a wider variety of other cosmological information. Our approach is also complementary to the many other analyses in the literature (e.g. \citealt{2007ApJ...666..716D}, \citealt{2008arXiv0804.4142K}, \citealt{2009arXiv0901.4804H}) that have 
used data from all possible sources.

In our approach we concentrate on the information from cosmological measurements that cover the same range of redshifts as the SDSS SN sample.  
Our aim is not to derive the most stringent limit on cosmological parameters available, but rather to verify that if we restrict ourself to probes coming from
a small and similar redshift slice the results on cosmological parameters remain stable and consistent.
This approach is warranted because of the growing emphasis on controlling systematic uncertainties in the analysis of cosmological data. There are clearly a number of systematic uncertainties that could affect the use of SNe as cosmological probes which likely depend on, or change with, redshift, including SN evolution (e.g., changes in the metallicity of progenitor stars \cite{2003ApJ...590L..83T, 2009ApJ...691..661H, 2009ApJ...693L..76S}), intergalactic dust \citep{2007ApJ...664L..13C, 2008MNRAS.386..475H}, Malmquist bias and the effects of gravitational lensing and peculiar velocities \citep{2006PhRvD..73l3526H}. 
Moreover, the photometric uncertainties associated with combining SN data from multiple surveys, over a range of redshifts, is already seen as the main limitation in using presently available datasets \citep[see][]{2009arXiv0901.4804H}. Our analysis addresses this issue by focusing exclusively on the SDSS SN sample, which is derived from a well-understood and stable photometric system. The SDSS has a  relative photometric accuracy of better than $2\%$ in griz, and  $4\%$ for the u--band (Padmanabhan et al. 2007), while the absolute calibration is also known to be of the order of $1\%$, leading to a homogeneous set of SN light-curves with high photometric accuracy \citep[see][]{2008AJ....136.2306H}. This set of data is robust to uncertainties in light-curve fitting. For example, the MLCS2k2 and SALT2 fits to the SDSS-only sample are shown to agree well  
\citep[see Section 10 in][]{kessler}
which is not the case when the higher redshift SN samples are added. 

The outline of this paper is as follows. In Section \ref{chapter_1st_year_data}, we describe the
SDSS-II SN data and use that data in Section 3 to study the cosmic acceleration in the Universe. Section 4 then compares the SDSS-II SN luminosity distances to the BAO distances from the SDSS and 2dFGRS, checking the distance duality relation. We then derive in Section 5 constraints on $w$
by combining the best-fit luminosity distances for SDSS-II SNe
with the growth rate of structure measurements taken from \citet{2003MNRAS.346...78H}
and a new measurement of the ISW effect taken from  \citet{2008PhRvD..77l3520G}. We conclude in Section \ref{conclusions}.

\begin{figure}
\center
 \includegraphics[width=80mm] {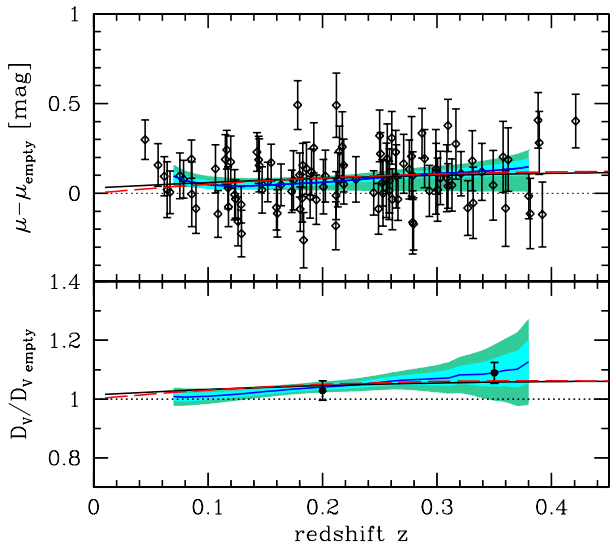}
 \caption{Residual Hubble diagram with respect to an empty universe for the 
103 Type Ia SNe from the first year of operation of the SDSS-II SN Survey. The red line 
shows a $\Lambda$CDM model with $(\Omega_M, \Omega_\Lambda) = (0.3, 0.7)$, similar to
our best-fit model given in Table \ref{table_w}. The blue line is the best fit to the data 
derived from the ``sliding window" technique described 
in Section \ref{sec:nonpa}, and the cyan and green shaded regions correspond to 
the $1\sigma$ and $2\sigma$ confidence intervals, respectively. The black line indicates 
an expansion history with a deceleration parameter of $q_0 = -0.34$ as described in 
Section \ref{section_global_fit}. The lower panel shows the same parameterization but now 
converted to $D_V$ according to Eq. (\ref{equation_dv}). The two data points represent 
the BAO measurements from \citet{2009arXiv0907.1660P}.}
 \label{fig_sdss_hubble}
\end{figure}

\section{The first-year SDSS SN data}
\label{chapter_1st_year_data}

The SDSS-II Supernova Survey \citep{2008AJ....135..338F} was part of the SDSS-II project and was focused on constructing a large sample of intermediate-redshift SNe 
($0.045<z<0.42$). One of the strengths of the SDSS-II SN Survey is that it builds upon the existing (and stable) infrastructure from the original SDSS 
(\citealt{1996AJ....111.1748F}; \citealt{1998AJ....116.3040G}; \citealt{2000AJ....120.1579Y}; \citealt{2001ASPC..238..269L}; \citealt{2001AJ....122.2129H}; 
\citealt{2001AJ....122.2267E}; \citealt{2002AJ....124.1810S}; \citealt{2003AJ....125.1559P}; \citealt{2006AJ....131.2332G}). The SDSS-II SN Survey is based on repeat 
imaging of ``Stripe 82", a region of the original SDSS with significantly deeper photometry than the regular SDSS survey. This region is $\sim120^{\circ}$ long 
and $2.5^{\circ}$ wide, centered along the celestial equator and extending from $20^{\rm hrs}$ to $4^{\rm hrs}$ in right ascension (passing through $0^{\rm hrs}$).

The SDSS-II SN Survey was carried out in three observing campaigns from September through December in each of 2005, 2006, and 2007 (there were also some observations for a short period in 2004). The 
new imaging data were initially reduced using the standard SDSS pipelines
\citep{2002AJ....123..485S, 2002AJ....123.2121S, 2004AN....325..583I, 2006AN....327..821T, 2008ApJ...674.1217P, 2008ApJS..175..297A}, followed by specific image-subtraction software to identify transient objects \citep{2008AJ....135..348S}. To determine the nature of these transients, the data were both visually inspected and fit with SN models. Subsequently, objects with a high probability of being a SN Ia \citep{2008AJ....135..348S} were spectroscopically observed using a variety of telescopes around the world \citep{2008AJ....135..338F, 2008AJ....135.1766Z}. 

In this paper, we only consider the 2005 observing campaign (the first
year of operation) as the data from other years is still being collated and analysed.
That year, the SDSS-II SN Survey discovered 130 spectroscopically confirmed type Ia supernovae and an additional 16 spectroscopically probable SNe Ia.
In \citet{kessler} distance moduli are obtained for 103 of the spectroscopically confirmed SNe Ia that pass stringent light-curve quality cuts, 
using the MLCS2k2 light-curve fitting routine. 
In the upper panel of Fig. \ref{fig_sdss_hubble} we show for these 103 SNe the residuals of the distance modules with respect to an empty universe.
We refer the reader to Section 9 of 
\citet{kessler} for an extensive discussion of systematic effects caused by changing the various light-curve fitting parameters.

We also direct the reader to Section 11 of \citet{kessler} for a comparison of the MLCS2k2 and SALT2 \citep{2007A&A...466...11G} light-curve fitters. They show that for the SDSS--only data the systematic difference between these two light-curve fitting methods is only $0.04$ in $w$, while Fig. 42 of their paper shows the two methods give consistent distance moduli for the same SDSS-II supernovae.
We also highlight that the two methods give similar contours in Fig. 26a and 35a of \citet{kessler} when comparing the full cosmological fits to the SDSS--only data.
This motivates the analysis in this paper and means our results are unaffected by the choice of light-curve fitter used. We restrict the analysis to the MLCS2k2 reduction taken from Table 10 in \citet{kessler},  as these data include corrections for selection effects.

When using these SDSS-II SNe for cosmological fitting, we calculate the confidence intervals via the $\chi^2$ statistic,

\begin{center}
\begin{equation}
\chi^2 = \sum^{N}_{i=1} \frac{(\mu_{{\rm LC},i}(z_i)-\mu_{\rm model}(z_i, \vec{x}))^2 }{ \sigma^2_{{\rm LC},i}+\sigma^2_{\rm sp}+\sigma^2_{\rm int} },
\label{equation_chi2}
\end{equation}
\end{center}

\noindent where $\mu_{\rm LC}$ and $\sigma_{\rm LC}$ are the distance moduli and errors, respectively, derived from the light-curve fitting method 
\citep[see][]{kessler}, 
and $\mu_{\rm model}$ are the expected distance moduli according to parameters $\vec{x}$ of the assumed cosmological model. Uncertainties in the measured spectroscopic 
redshift and peculiar velocity of the SN are taken into account using

\begin{equation}
\sigma_{\rm sp} =  \frac{5}{\ln(10)} \frac{(1+z)}{z(1+z/2)} \sqrt{(\Delta z)^2+(\Delta v_{\rm p})^2/c^2},
\label{eq_delta_z}
\end{equation}

\noindent where $\Delta z$ is the measurement uncertainty in redshift and $\Delta v_{\rm p}$ is the characteristic amplitude
of the peculiar velocities, which we take to be $360~\mathrm{km~s^{-1}}$ (see, e.g., \citealt{2006ApJ...653..861M}). 
We investigated the effect of correlated peculiar velocities of SDSS-II SNe and found no detectable effect (see Appendix B).
We therefore do not include peculiar velocity correlations in further analyses. Besides peculiar velocities we ignore 
further potential correlations between individual SNe and treat them according Eq. (\ref{equation_chi2}) as statistical independent.

Following standard practice, we add an intrinsic dispersion in distance modulus, $\sigma_{\rm int}^2$, in the 
denominator of Eq. (1) and determine it by setting $\chi^2/N_{\rm dof}=1$ for the best-fit cosmological model.  
This term accounts for the fact that the errors on the distance moduls reported by the light curve fitter could underestimate
the real error if we assume that a smooth cosmology is the correct underlying model.
For the best-fit model with constant $w$, we find $\sigma_{\rm int}=0.088$ 
mag for our SDSS-only SN sample, while in \citet{kessler} a value of $\sigma_{\rm int}=0.16$ derived from the nearby
supernova samples is used in combination with the SDSS data.
This results in broader contours compared to the one shown in this paper (see their Fig. 26a, and Appendix E for a possible explanation), 
but only marginally changes the most likely values of $w$ and $\Omega_M$. Similarly if we omit any intrinsic dispersion
we get narrower contours but only slight changes in $w$ and $\Omega_M$ well within the errors given in Table \ref{table_w}.

A further complication arises due to the uncertain calibration of the absolute magnitude (at peak) of SNe~Ia,
leading to a degeneracy with the absolute value of $H_0$. To account
for this, we marginalize over $H_0$. This procedure makes use of the relative distances reported by
the light curve fitter but not of their absolute value. In a recent paper \citet{2009ApJ...699..539R} redetermined
the Hubble constant $H_0$ or equivalently the absolute brightness of SN Ia. One could use this value obtained from 
measurements in the nearby universe as a prior on $H_0$ in our analysis.
But - as laid out in the introduction - we want to limit our analysis to probes taken at comparable redshifts as the SDSS SNe
and therefore refrain from using this additional information.
 
\section{Testing Cosmic Acceleration}
\label{section_global_fit}

Given the homogeneity of the SDSS-II SN dataset, we begin our analysis by revisiting the original evidence for cosmic acceleration in the expansion of the universe. In detail, if we simply assume the universe is homogeneous and isotropic, and is described by a Robertson-Walker 
metric with a scale factor of $a(t)$, then the purely kinematic deceleration parameter, $q$, is defined by

\begin{equation}
q(z) \equiv -\frac{\ddot a a}{\dot a^2},
\end{equation}

\noindent where $q<0$ corresponds to acceleration (we also assume light follows null geodesics and is therefore redshifted in the usual way).  We relate $q$ to the Hubble parameter by

\begin{equation}
H(z)\equiv\frac{\dot a}{a}=H_0 \; \exp\left(\int_0^z \frac{1+q(\zprime)}{1+\zprime} d\zprime\right).
\end{equation}
Luminosity distance can then be calculated directly from the expansion history via
\begin{equation}
d_L(z) = c(1+z)\int_0^z \frac{d\zprime}{H(\zprime)} ,
\end{equation}
where we have assumed a flat universe.  Thus, the magnitudes and redshifts of any SNe can be used to constrain $q(z)$ without choosing a particular dark-energy model, or even a particular theory of gravity. The assumption of flatness is necessary in practice since constraints on $q(z)$ degrade significantly when curvature is allowed to vary.  A prior on curvature from CMB measurements would, of course, strengthen the constraint, but such a prior is based on the validity of GR, counter to the intention of the $q(z)$ analysis.

The simplest deceleration model we can fit is a constant, $q(z)=q_0$.  In this case, the luminosity distance simplifies to
\begin{equation}
d_L(z) = \frac{c(1+z)}{H_0q_0}\left[1-(1+z)^{-q_0}\right].
\label{eq:LumDistq0}
\end{equation}
\noindent
Fitting Eq. (\ref{eq:LumDistq0}) to the SDSS-only SN data, we find a best fit of $q_0 = -0.34$ and $h = 0.636$. Marginalizing the joint 
probability density function (PDF) over $h$, we find $q_0=-0.34\pm0.18$, or $q_0<0$ with 97\% probability, i.e., the SDSS alone finds evidence for acceleration 
at 2$\sigma$ without concerns regarding the absolute calibration of the peak brightness of SNIa and the relative calibration between SN surveys.

\begin{figure}
\begin{center}
\includegraphics[width=80mm]{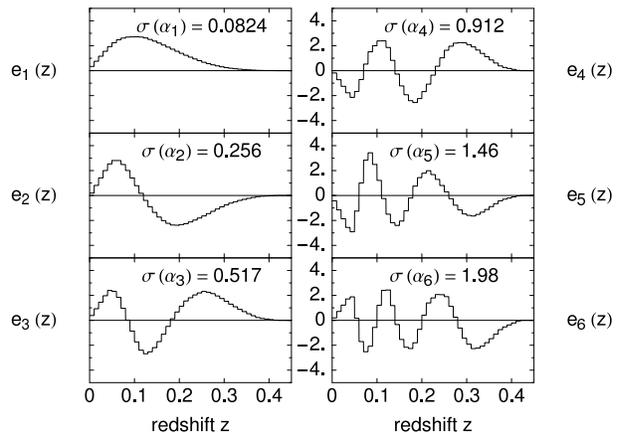}
\end{center}
\caption[First mode]{The principal components for $q(z)$ derived from the SDSS-II SN data.  Each mode includes a Fisher matrix estimation of the error bar, $\sigma(\alpha_i)$, for its coefficient $\alpha_i$.  The errors are uncorrelated, and we have ordered the modes according to the size of their error bars.  Only the first six modes are shown.
\label{fig:6modes}  }
\end{figure}

There is no reason to expect $q(z)$ to be constant; indeed, under \LCDM, $q$ evolves from 1/2 during matter 
domination to $-1$ when vacuum energy dominates.  However, we find that including additional parameters in $q(z)$ (e.g. by Taylor expanding $q$) degrades our constraints to the point of being uninteresting.

\subsection{Principal Component Analysis}

When trying to reconstruct an unknown function from noisy data, there is the concern that particular features of the reconstruction are not indicative of the true underlying function, but an artifact of the chosen parameterization.  This concern is magnified for a function like $q(z)$, which is related to the data $\mu(z)$ by two derivatives.  Therefore, we need a more robust way to determine if the universe has accelerated --- a way that does not depend on the naive assumption of a constant $q(z)$.  A principal-component analysis \citep{2003PhRvL..90c1301H} can be used to address this issue in a parameter-independent way.  Principal components are a unique set of orthogonal basis functions ($e_i(z)$), such that
\begin{equation}
q(z) = \sum_i \alpha_i e_i(z),
\label{eq:PCdecomp}
\end{equation}
which allows us to specify $q(z)$ using the coefficients, $\alpha_i$.  The principal components, or ``modes''  $e_i(z)$, are explicitly constructed so that the coefficients can be measured independently of each other, i.e., they have uncorrelated error bars.  To construct these modes, we start with a piecewise-constant parameterization of $q(z)$ in bins of $dz=0.01$, and we use our data to calculate a Fisher matrix for this parameter set and the Hubble parameter $H_0$.  After marginalizing over $H_0$, the modes $e_i(z)$ are given by the eigenvectors of the resulting matrix.  We are free to normalize these functions so that
\begin{equation}
\int e_i(z) e_j(z) \; dz = \delta_i^j, 
\end{equation}
which now specifies each function up to an overall sign convention.
This procedure is completely general; in the limit of $dz\rightarrow 0$, we can specify any continuous function using these modes.  The procedure is parameter-independent in the sense that we do not specify the modes a priori: they are determined primarily by the data.  We do choose $q(z)=0$ as the fiducial model for our Fisher matrix calculation, but the resulting modes are insensitive to this choice.

The six modes constrained best by the SDSS-only SN data are shown in Fig. \ref{fig:6modes}.  Each mode is accompanied by an estimate of the error bar on its coefficient, $\sigma(\alpha_i)$, calculated from the eigenvalues of our Fisher matrix.  The first coefficient, $\alpha_1$, has the tightest constraint, while higher mode coefficients have increasingly larger errors and so provide little information about the shape of $q(z)$.  Note that the first mode does not cross zero, and we have chosen its sign to be wholly positive. As argued by \citet{2006ApJ...649..563S}, this mode is useful since in addition to being well constrained, $\alpha_1$ can only be negative if $q(z)$ is negative for some $z$.  Therefore, if we fit the model $q(z)=\alpha_1 e_1(z)$ to the data, and then measure $\alpha_1<0$, it constitutes parameter-independent evidence that the universe has accelerated at some redshift.  We need not consider additional modes since the constraint on $\alpha_1$ is, by construction, independent of the constraints on the other $\alpha_i$.  Marginalizing over $h$, we find $\alpha_1=-0.155\pm 0.086$ and determine that $\alpha_1<0$ with 96\% probability.  The fact that the error bar from our fit closely matches the Fisher matrix estimate of $\sigma(\alpha_1)=0.082$ demonstrates consistency.   This result is comparable to our $q_0$ fit above, while providing robust evidence for an accelerating universe regardless of the model assumptions. 

The significance of our detection could be enhanced by combining the SDSS-II SNe with other SN datasets. This has already been done in part by \citet{2006ApJ...649..563S} but could suffer from systematic uncertainties associated with combining data from different instruments and surveys. Again, the reader is referred to \citet{kessler} for a detailed description of such combinations of 
the SDSS-II SN Survey with other SN datasets. We plan to repeat this analysis with the full 3--year SDSS-II SN dataset.

\section{Comparison of Distances}

We next consider the comparison of our SDSS-II SN distances with other geometrical distance estimates over the same redshift range. This is motivated by the original results of \citet{2007MNRAS.381.1053P} who noticed some tension (at $>2\sigma$) between the cosmological constraints derived from nearby BAO measurements ($z=0.2$ and $z=0.35$) and higher redshift SNe of Astier et al. (2006). The
BAO provide a measure of distances in the universe by relating the scale of the sound horizon at last 
scattering ($r_s$) to the scale of the corresponding correlations seen in the galaxy distribution. One such measurement of this ratio is given by the 
$A$-parameter in \citet{2005ApJ...633..560E}. This parameter is frequently used in combination with SN data to derive constraints on $w$ \citep{2006A&A...447...31A, 2008arXiv0804.4142K}, e.g., see  \citet{kessler} for the combination of the SDSS-II SN data with the $A$-parameter. 

Here, we adopt the quantity 
\begin{equation}
D_V(z) = \left[D^2_M \frac{cz}{H(z)}\right]^\frac{1}{3},
\label{equation_dv}
\end{equation}
\noindent defined in \citet{2005ApJ...633..560E} where $D_M$ is the co-moving distance. \citet{2007MNRAS.381.1053P} showed by combining measurements of $r_s/D_V$ at redshifts of 0.2 and 0.35 from both the 2dFGRS and SDSS galaxy samples, that one can obtain the ratio of the distance between 
two different redshifts that is independent of both $r_s$ and $H_0$. This approach also avoids a large extrapolation between the redshift of 
recombination ($z_{\rm CMB}=1090$, \citealt{2008arXiv0803.0547K}) and these intermediate-redshift measurements. 

For the analysis presented in this paper, we have adopted the latest value of $D_V(z=0.35)/D_V(z=0.2) = 1.736\pm0.065$ taken from \citet{2009arXiv0907.1660P}.
The inferred value of $D_V(z=0.35)/D_V(0.2)$ from Percival et al. (2009) is lower than that of Percival et al. (2007), bringing it into better agreement with $\Lambda$CDM.
This change was caused by a revised error analysis and a change in the methodology adopted, as well as the addition of more data. In this paper we do not use constraints on $r_s(z_d)/D_V(z)$, which depend on the sound horizon at the baryon drag epoch $r_s(z_d)$. We therefore avoid including CMB data, commonly used to model the sound horizon.

\subsection{The Sliding Window Method}
\label{sec:nonpa}

In Fig. \ref{fig_sdss_hubble} we show the Hubble diagram for the SDSS-II SN data discussed in Section 2 compared with a variety of cosmological and non--parametric models discussed herein. We find a scatter of $0.14$ mag around these models independent of the particular fitting method. The first non-parametric model we consider is a ``sliding window" method which allows us to investigate the general shape and smoothness of the Hubble diagram without assuming a cosmological model. We have thus fit piecewise Hubble parameters and luminosity 
distances in different redshift bins using a local redshift window following the approach of \citet{2003ApJ...597....9D}. At each redshift $z$, we fit the SNe 
co-moving distances, $\DM(z_i)=\DL/(1+z_i)$, over a fitting window of $z_i-\Delta z <z_i< z_i+\Delta z$ 
\footnote{We actually allow for a tapered region at both ends of this redshift window where 
the weights (i.e., the inverse-squared errors) are reduced by a Gaussian function with a standard deviation of $\sig_z=0.02$. 
This leads  to a suppression of fluctuations due to the inclusion of individual data points into the window (see \citealt{2003ApJ...597....9D}).}, by a 
polynomial of second order given by:
\begin{equation}
f(z_i-z)=A_0+A_1 (z_i-z)+A_2 (z_i-z)^2.  
\label{eq_sliding_window}
\end{equation}
 The values of $A_i$ are determined separately via $\chi^2$-minimisation in each redshift window, and we slide this window as a function of redshift in increments of 0.01 throughout 
the entire range. The best-fit $\DM$ at $z$ is proportional to $A_0$, while the best-fit $c/H$ is proportional to $A_1$, and $D_V$ is related to $[ zA^2_0 A_1] ^{1/3}$. Our results depend on the size of the redshift window, with a wider window allowing less flexibility but smaller errors, and vice versa.

We show in Fig.\ref{fig_sdss_hubble} the resulting non--parametric fit to the SDSS-II SN data as an example for a window size of $\Delta z = 0.15$ which 
demonstrates that the SDSS SNe data is fully consistent with the individual BAO measurements at $z=0.2$ and $0.35$ of \citet{2009arXiv0907.1660P} (the cyan and green shaded regions indicate the 1$\sigma$ and 2$\sigma$ errors, which are highly correlated as they share the same data points in the overlapping fitting windows). 
Next, we derive the ratio 
$\DV(0.35)/\DV(0.2)$ and determine the covariances between the $A_i$ values within, and between, redshift bins using the observational errors. This is shown in Fig. \ref{fig_bao} for different window sizes. It is interesting to note that the sliding window method tends to prefer large values (steeper slopes) for $\DV$ compared to 
$\DV$ calculated from a $q_0$ fit described in Section 3, or $\DV$ from the best fitting $w-\Om$ parametrizations given Table 1. We also see that the sliding window method provides values for $\DV$ which are fully consistent with the BAO result. 
We also show in Fig. \ref{fig_bao} the best-fit ratios of $\DV$ when the SDSS-II SN data are simultaneously fit with the BAO data. The SN constraints dominate these 
results because of their smaller uncertainty.

\subsection{Testing the Distance Duality Relation}
\label{duality_test}

\begin{figure}
\includegraphics[width=80mm] {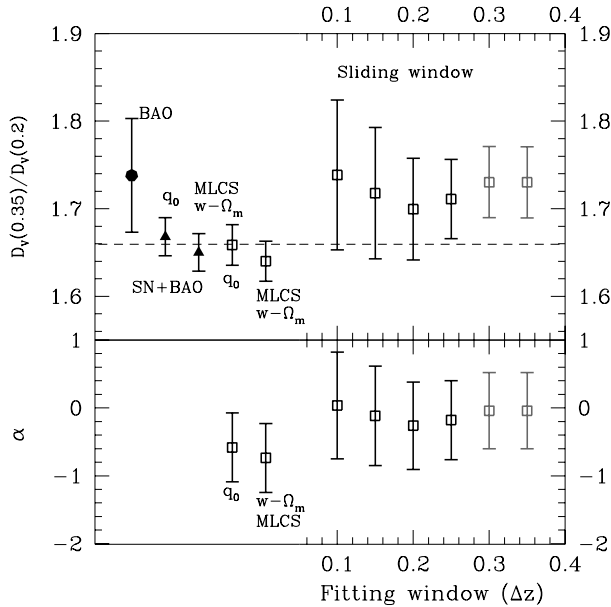}
 \caption{Upper panel: Measurements of the ratio of $D_V(0.35)/D_V(0.2)$ using different fits to the SN data, in comparison to or in combination with the BAO 
(solid circle) measurement. The points labelled $q_0$ indicate $D_V$ derived using the fit in Section \ref{section_global_fit}. Points labelled $w-\Omega_M$ are derived from the $w-\Omega_M$ parameterization marginalized over $0<\Omega_M<1$ and $-2<w<0$. Points shown as triangles include the BAO measurement as a prior whereas boxes
are without the BAO. The points to the right show the values derived from the ``sliding window" method as a function
of the window width. The grey points are for values of the sliding window that are comparable in size to the redshift range of the whole SDSS data-set and therefore, approach the global fit.
The dashed line indicates $\Lambda$CDM with $\Omega_M=0.3$ and $\Omega_{\Lambda}=0.7$. 
Lower panel: Given the above-mentioned parameterizations, we show the best-fit value
of $\alpha$ defined in Eq. (\ref{equation_duality_test}).}
\label{fig_bao}
\end{figure}

In the following we use the famous reciprocity relation \citep{1933PMag...15..761E, 1971grc..conf..104E}, or distance duality
to compare the SN and BAO distance scales. In detail, the angular diameter distance and luminosity distance are related by

\begin{equation}
\frac{d_L}{d_A}=(1+z)^2 
\label{equ_distance_duality}
\end{equation}
(for a discussion, see e.g, \citealt{2004PhRvD..69j1305B}). This relation relies on photon conservation, but holds for any
geometry and any metric theory of gravity where photons follow null
geodesics. Therefore, it is a general test of our underlying assumptions about the nature of our Universe.

One might have expected that the distance duality relation has already been 
tightly constrained by observations of the blackbody CMB spectrum from the COBE FIRAS
experiment \citep{1994ApJ...420..439M}. However, this observation does not constrain 
deviations from distance duality as the photon
number may not be conserved (either through production or loss of
photons) or more radically, photons may not follow null geodesics. Also, a
grey dust component that absorbed photons independent of frequency
would not cause spectral distortions away from a blackbody in the CMB
since all frequencies would be affected equally. However, this grey
dust would cause strong deviations from distance duality since it
would make the luminosity distance to any objects larger while leaving
the angular diameter distance unchanged. Another way to hide the
distance duality effects from CMB observations would be to affect
photon number only at much higher or lower frequencies
than the microwave. This is, for example, what was needed to make the
axion-photon mixing proposal for the dimming of the SNe~Ia consistent
with CMB constraints \citep{2002PhRvL..88p1302C}.

Beyond the CMB, several other analyses, using similar data to that discussed herein, have reported evidence for violations of distance duality at the $\sim 2\sigma$ level \citep{2004PhRvD..69j1305B, 2008JCAP...07..012L}. We revisit this issue here using a methodology similar to that outlined by \citet{2009ApJ...696.1727M} and \citet{2009JCAP...06..012A} that does not rely on the absolute calibration of the distances to compute the ratio
$d_L(z)/(d_A(z)(1+z)^2)$. Instead, we check the relative behaviour of this
ratio as a function of redshift by testing the consistency of the ratio
at two redshifts, $z=0.2$ and $z=0.35$ where we now have updated BAO 
measurements from \citet{2009arXiv0907.1660P}.

In the following we parameterize the distance duality relation in what we call the $\alpha$-model as

\begin{equation}
d_L  = (1+z)^{2+2\alpha}d_A = (1+z)^{(1+\alpha)}D_M,
\label{equation_duality}
\end{equation}

\noindent where $\alpha=0$ represents the expected distance duality relation, and therefore $\alpha \ne 0$ 
indicates a possible violation. 
To quantify the discrepancy between the two measures, we replace $D_M$ in Eq. (\ref{equation_dv})
with $d_L$ from Eq. (\ref{equation_duality}), and derive the relation

\begin{equation}
\frac{(1+z_2)^{2+2\alpha}}{(1+z_1)^{2+2\alpha}}\frac{z_1}{z_2} = \frac{D_V^3(z_1)}{D_V^3(z_2)}
\frac{d_L^2(z_2)}{d_L^2(z_1)}
\frac{H(z_1)}{H(z_2)}.
\label{equation_duality_test}
\end{equation}

\noindent
On the right-hand side, $D_V(z_1)/D_V(z_2)$ is given by the BAO measurements, while $d_L^2(z_2)/d_L^2(z_1)$
and $H(z_1)/H(z_2)$ have to be inferred from the SDSS-II SN data.
We calculate $d_L$ and $H(z)$ using
the two methods introduced above. First, we use the results from the $q_0$ fitting,
and subsequently we calculate $d_L(z, q_0)$ and $H(z, q_0)$ at 
redshifts of $0.2$ and $0.35$. The parameter $\alpha$, and its error, are then calculated by Eq. (\ref{equation_duality_test}).
We note that Eq. \ref{equation_duality_test} probes the consistency of the ratio at one redshift given the ratio at the other redshift and
is therefore not sensitive to any scaling proportional to $(1+z)^2$ but would be sensitive to any other loss function.

For the second method, we use the sliding window technique to derive $\vec{D}_{V,\;{\rm SN}} \equiv [zA^2_0 A_1] ^{1/3}$ (see Eq. (\ref{eq_sliding_window})) 
and the corresponding covariance matrix at the two redshifts ($z=0.2$ and $z=0.35$)
where we have BAO measurements. 
The best fit and error of $\alpha$ is calculated by applying Bayes theorem. 
In detail we model $\vec{D}_{V,\;{\rm BAO}}=\beta (1+z)^{2\alpha/3}\vec{D}_{V,\;{\rm SN}}$ at the two redshifts
based on Eq. \ref{equation_duality_test}
where $\beta$ is a free scale parameter absorbing $H_0$ and the scale of the sound horizon $r_s$ at recombination.
We then calculate the likelihood of the BAO $\vec{D}_{V}$ measurements for $\alpha$ and $\beta$, by integrating
over all possible SN $\vec{D}_{V}$ at $z=0.2$ and $z=0.35$ given the gaussian prior $p(\vec{D'}_{V}|\vec{D}_{V,\;{\rm SN}})$
constructed from the results of the sliding window technique:

\begin{eqnarray}
L(\vec{D}_{V,\;{\rm BAO}}|\beta,\al) = \int d\vec{D'}_{V} \times \hspace{3.5cm} \nonumber \\
 \hspace{0.9cm} \left[ L(\vec{D}_{V,\;{\rm BAO}}|\beta(1+z)^{2\al/3}\vec{D'}_{V})\;p(\vec{D'}_{V}|\vec{D}_{V,\;{\rm SN}}) \right],
\end{eqnarray}
The covariance matrix used in $L(\vec{D}_{V,\;{\rm BAO}}|\beta(1+z)^{2\al/3}\vec{D'}_{V})$ is taken from Percival et al. (2009).
Applying this procedure and subsequently marginalizing over $\beta$, we calculate the best fit $\alpha$ and its error. The results 
are shown in the lower panel of Fig. \ref{fig_bao}, along with the result from the $q_0$ method. 

For the $q_0$ parameterization, we find  $\alpha=-0.55\pm0.45$, while for the sliding window scheme, we find that all results agree with $\alpha=0$ within one sigma, i.e., the errors on $\alpha$ are  $\simeq 0.5$ for most of the window sizes shown in Fig. \ref{fig_bao}. 
These results re-enforce our findings that the SN and BAO distance scales are now in good agreement over this redshift range (confirming the new findings of \citet{2009arXiv0907.1660P}).

\begin{figure}
\includegraphics[width=80mm] {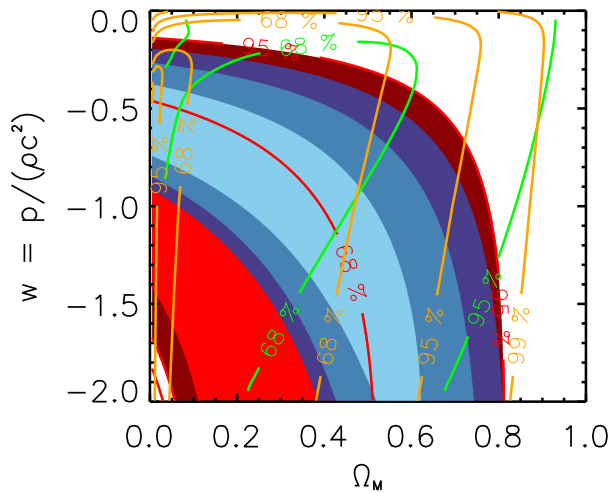}
\caption{Confidence intervals (68\%, 95\%, and 99\%) in the $w-\Omega_M$ plane
for a flat universe and a $w$CDM model derived from the SDSS SNe (shaded blue region), the
BAO $D_V(0.35)/D_V(0.2)$ ratio (filled red, upper 99\% contours are off the plot), redshift-space distortions (orange), 
and the ISW effect (green).
In regions where the SN contours overlap the BAO contours the latter are indicated as red lines.}
\label{fig_sn_bao_qs_isw}
\end{figure}

\section{Constraining Cosmological Parameters}

In contrast to the previous sections, which focused on kinematic
constraints of the cosmic expansion, here
we investigate the constraints on standard cosmological parameters using the SDSS SN data only in combination with dynamical measurements,
eg. from the BAO, redshift space distortions and ISW.
This will have less statistical power than using the combination of SN datasets presented in \citet{kessler} and \citet{davis2008} but our analysis is complementary to these companion papers and maximizes the impact of the homogeneity of the SDSS data (both for the SN and BAO). 

In Fig. \ref{fig_sn_bao_qs_isw}, we begin by showing our constraints on $w$ and $\Omega_M$ using only the SDSS SN data. This is achieved using  the $\chi^2$ statistic according to Eq. (\ref{equation_chi2}) over a three dimensional grid of $200$ bins in $w$, $100$ bins in $\Omega_M$ and $160$ bins in $H_0$ ranging from $40$ to $80~\rm{km~s^{-1} Mpc^{-1}}$.
Subsequently, we convert the $\chi^2$ into a 
likelihood using $L= \exp(-\frac{1}{2}(\chi^2 - \chi^2_{\rm min}))$, where $\chi^2_{\rm min}$
is the minimum $\chi^2$ found in the parameter space and, in our case, is
by definition close to the number of SNe in the dataset (as we have added $\sigma_{\rm int}$).
We then marginalize over $H_0$ by summing the likelihoods over the $H_0$-bins.
The shaded blue contours show the resulting 
confidence levels (68\%, 95\%, and 99\%) in the $w-\Omega_M$ plane under the 
assumption of a flat universe ($\Omega_{\Lambda} = 1-\Omega_M$). 
These contours include statistical errors only; systematic 
errors are discussed in Section \ref{chapter_systematic}.

\begin{figure}
\includegraphics[width=80mm] {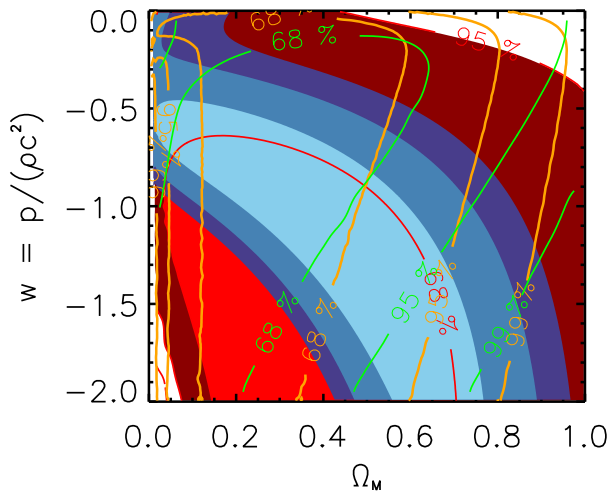}
\caption{Similar to Fig. \ref{fig_sn_bao_qs_isw}, but now including
curvature as described in Appendix \ref{appendix_curvature}.}
\label{fig_sn_gs_isw_curvature}
\end{figure}

Fig. \ref{fig_sn_gs_isw_curvature} is similar, but here - instead of assuming a flat universe - we allow for curvature 
according to the CMB shift parameter $R$. We do this by
by calculating for a given ($w$, $\Omega_M$) combination the corresponding value of $\Omega_{\Lambda}$
according to the constraints from $R$ (see Appendix \ref{appendix_curvature} 
for more details on the treatment of curvature). As discussed in Section 1, we see little effect on our results from allowing curvature to vary because of the  
relative small deviations from flatness allowed by the WMAP data.
For results see Table 1. 

For comparison with the blue SN contours, we also provide in Fig. \ref{fig_sn_bao_qs_isw} and Fig. \ref{fig_sn_gs_isw_curvature} the (red) contours for the BAO measurements from 
\citet{2009arXiv0907.1660P}. The BAO measurements are in reasonable agreement with the SDSS--only SN contours but still prefer $w<-1$ as originally discussed in \citet{2007MNRAS.381.1053P}, although all measurements are consistent with a cosmological constant. As the BAO and SN distances have reasonable overlap, 
we provide in Table \ref{table_w} various constraints on $w$ and $\Omega_M$ derived from combinations of the SN data with intermediate redshift dynamical measurements 
described in the following Sections and the BAO. We note that the SN data dominates the width and position in the $w$-direction of the likelihood contours, as they have smaller 
errors compared to the BAO constraints, while adding the BAO data helps to curtail the  large values of $\Omega_M$ seen for the SDSS--only results.

\subsection{Constraints from dynamical measurements}
\label{chapter_dynamic}

To improve the cosmological constraints, we can include a number of other low redshift cosmological measurements thus providing a first measurement of the 
cosmological model within the local universe. In particular, we consider constraints on $w$ derived from measurements of the growth of 
structure in the universe, including redshift-space distortions from the 2dFGRS and the ISW effect from the SDSS imaging survey. Both of these methods are particularly sensitive to $\Omega_M$ 
and thus provide independent and orthogonal constraints to the SN data. 

\subsubsection{Redshift-Space Distortions}
\label{section_gs_structure}

Under the assumption that galaxies are related to the large scale dark-matter
distribution, the anisotropy of the redshift-space correlation
function depends on the parameter
\begin{equation}
  \beta(z) = f_g(z)/b(z),
\end{equation}
where $b(z)$ is the linear bias relating the galaxies to the underlying dark
matter and $f_g(z)$ the growth rate of structure. Absolute deviations
between the real-space and redshift-space correlation functions depend
on the parameter combination $f_g(z)\sigma_8(z)$, where $\sigma_8(z)$
is defined as the root-mean square (rms) mass fluctuation in spheres of 
radius 8\,h$^{-1}$ Mpc, and provides a convenient way of normalising the 
matter fluctuations (for a recent review see \citealt{2008arXiv0808.0003P}).

\begin{figure}
\includegraphics[width=80mm] {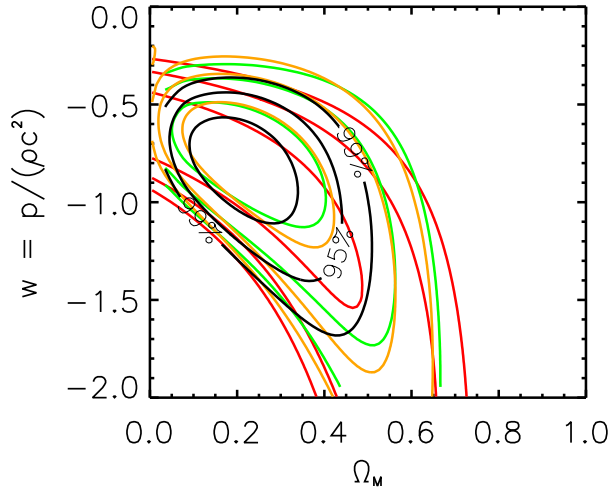}
 \caption{Confidence intervals (68\%, 95\%, and 99\%) in the
$w-\Omega_M$ plane for the combination of SDSS-II SNe with BAO (red), redshift-space distortions (orange), 
ISW effect (green) and for the combination of SN, BAO, RS and ISW (black) under the assumption of a flat universe. Numerical results are given in Table \ref{table_w}.}

 \label{fig_sdss_plus_gs_and_isw}
\vspace{3mm}
\end{figure}

To remove either the dependence on galaxy bias from a measurement of
$\beta(z)$, or equivalently the dependence on $\sigma_8(z)$ from the
measurement of $f_g(z)\sigma_8(z)$, we need further
cosmological information. 
In this paper, we adopt the central value of
$\beta(z\approx 0.15)=0.49\pm0.09$ calculated by \citet{2003MNRAS.346...78H} from the 2dFGRS, 
which is an update of the measurement given by \citet{2001Natur.410..169P}. We follow the
procedure outlined by \citet{2008Natur.451..541G}, and convert from $\beta$ to $f_g$ by adding an
additional uncertainty of $0.12$ in quadrature to account for the
uncertainty in galaxy bias, which is estimated to be close to
unity. This error includes the cosmological dependence of the bias
measurement. For further calculations, we thus use $f_g(z=0.15) =
0.49\pm0.15$, assuming the weighted median redshift of the 2dFGRS
survey of $z\approx0.15$.

We adopt the parameterization given by \citet{2005PhRvD..72d3529L},

\begin{equation}
g(a) = \frac{\delta}{a} = \exp \int_0^a d \ln a [\Omega_M (a)^\gamma -1],
\label{ga_linder}
\end{equation}

\noindent which is related to the growth rate by
\begin{equation}
f_g = \frac{\dot{\delta}}{g},
\label{eq_growthrate}
\end{equation}

\noindent where $\delta = \delta \rho_M/\rho_M$ describes the perturbations in the
density of matter ($\rho_m$). For constant $w>-1$, the exponent $\gamma$ in Eq. (\ref{ga_linder}) can be
approximated as

\begin{equation}
\gamma = 0.55 +0.05(1 + w),
\end{equation}

\noindent while for a phantom-like dark energy component (with $w<-1$) the exponent is 

\begin{equation}
\gamma = 0.55 +0.02(1 + w).
\end{equation}

Solving
Eq. (\ref{eq_growthrate}) numerically, we derive the (orange) contours shown in 
Fig. \ref{fig_sn_bao_qs_isw} and Fig. \ref{fig_sn_gs_isw_curvature}. In Table \ref{table_w}, we present the constraints on $w$ from the 
combinations of these data with the SDSS-II SN and BAO likelihoods, marginalized over $\Omega_M$ and vice versa.

\begin{table}
  \caption{
  Results for $w$ and $\Omega_M$ from combinations of the SDSS-II SN and BAO data with redshift-space distortions (RS) 
and the ISW effect. We provide measurements based on the MLCS2k2 light-curve fitting method and our assumptions about 
curvature (Geo.). Values for $w$ are derived after marginalizing over $\Omega_M$ and vice versa. The given uncertainties are 
statistical errors only; for an estimate of the systematic uncertainty see Section \ref{chapter_systematic}. 
}
\begin{tabular}{@{}ll cc@{}}
\hline
Dataset  & Geo. & $w$ & $\Omega_M$ \\
\hline
SN+BAO                & flat & $-0.74^{+0.17}_{-0.32}$ & $0.37^{+0.16}_{-0.64} $\\
SN+RS                   & flat & $-0.77^{+0.19}_{-0.25}$ & $0.26^{+0.15}_{-0.12}$ \\
SN+ISW                 & flat & $-0.74^{+0.16}_{-0.22}$ & $0.23^{+0.15}_{-0.12}$ \\
SN+RS+ISW          & flat & $-0.76^{+0.17}_{-0.19}$ & $0.23^{+0.10}_{-0.08}$ \\
SN+RS+ISW+BAO & flat & $-0.81^{+0.16}_{-0.18}$ & $0.22^{+0.09}_{-0.08}$ \\
\hline
SN+BAO                & curved &  $-0.99^{+0.30}_{-0.59}$ & $0.50^{+0.13}_{-0.22}$ \\
SN+RS                   & curved &  $-0.82^{+0.19}_{-0.26}$ & $0.31^{+0.15}_{-0.12}$ \\
SN+ISW                 & curved &  $-0.78^{+0.16}_{-0.22}$ & $0.27^{+0.15}_{-0.13}$ \\
SN+RS+ISW          & curved &  $-0.79^{+0.16}_{-0.20}$ & $0.27^{+0.10}_{-0.09}$ \\
SN+RS+ISW+BAO & curved &  $-0.85^{+0.17}_{-0.19}$ & $0.27^{+0.10}_{-0.09}$ \\

\hline
\end{tabular}
\label{table_w}
\end{table}

\subsubsection{The Late-Time integrated Sachs-Wolfe Effect}

The integrated Sachs-Wolfe (ISW) effect is caused by the change in energy of CMB photons as they pass through a time-varying gravitational 
potential \citep{1967ApJ...147...73S}. In a flat, matter-dominated universe, we would not expect to see an ISW effect as the large-scale 
gravitational potentials do not change {\nbf in conformal time}. However, in a universe dominated by DE or curvature, we should detect a
so called {\it late-time} ISW effect, which provides 
a direct measure of these quantities at the redshift of the changing potentials, i.e., the effect does not depend on the previous history of 
the growth of structure. 

The late-time ISW effect introduces additional secondary anisotropies on top of the primary CMB fluctuations and is therefore hard to detect 
directly. However, the ISW effect can be seen via the cross-correlation of the CMB with tracers in the large-scale structure of the universe 
as outlined by  \citet{1996PhRvL..76..575C}. This has now been achieved by a number of authors using a host of different galaxy datasets 
\citep{2003ApJ...597L..89F,
2003astro.ph..7335S,
2004Natur.427...45B,
2004ApJ...608...10N,
2004PhRvD..69h3524A, 
2004MNRAS.350L..37F,
2005PhRvD..72d3525P,
2006MNRAS.372L..23C, 
2006PhRvD..74f3520G,
2007MNRAS.377.1085R, 
2008PhRvD..78d3519H,
2008ApJ...683L..99G}.

In this paper, we exploit the recent analysis of \citet{2008PhRvD..77l3520G} and focus on the subset of intermediate-redshift ($z<0.4$) SDSS data they used. 
Even this subset of data shows a detection of the ISW effect at the $3\sigma$ level \citep{giannantonio_private}. The contours for this new 
determination of the ISW effect are plotted in Figs. \ref{fig_sn_bao_qs_isw},  \ref{fig_sn_gs_isw_curvature}, and  \ref{fig_sdss_plus_gs_and_isw}. In Table \ref{table_w}, we present the combination of this new ISW effect measurement with our SDSS-II SN and BAO data, using the same procedure as discussed in Section \ref{section_gs_structure}.

\subsection{Evaluation of SN Systematics}
\label{chapter_systematic}

We provide in  Table \ref{table_w} measurements of $w$ and $\Omega_M$ from various combinations of the four data-sets considered herein (SN, BAO, RS, ISW). The most stringent constraint comes from the combination of all the probes giving $w=-0.81^{+0.16}_{-0.18}$ and $\Omega_M=0.22^{+0.09}_{-0.08}$, which is competitive given the restricted redshift range considered in this analysis. However, much of this constraint comes from the combination of just the SDSS-II SNe and ISW measurements (See Table \ref{table_w}). The ISW contours already include correlations between different angular and redshift bins and cosmic variance, 
and could therefore be considered stable (see e.g. \citealt{2008PhRvD..77l3520G}). Similarly the contours we use for redshift-space distortions include the dominant uncertainty coming from the galaxy bias (see the procedure laid out in Section \ref{section_gs_structure}). 
\cite{2009arXiv0907.1660P} has done several checks and found that their result is robust against variations in sample selection, number of redshift slices, calibration and other potential influences. 
Therefore, the results presented in Table 1 includes major uncertainties affecting the other probes but only the statistical uncertainties from the SDSS-SN data on the measured cosmological parameters. 

As discussed in  \citet{kessler}, the SDSS-II SN distances also depend on the detailed choices and assumptions within the MLCS2k2 supernova light-curve fitting procedure, including different training vectors, priors on $A_V$ and $R_V$, uncertainties in zero points and the filter systems, and selection biases. To quantify the systematic uncertainties associated with these parameter choices, we repeat our analysis above for these different choices and, following the procedure laid out in \citet{kessler}, we calculate a variation of $\Delta w=\pm 0.15$ with respect to the fiducial MLCS2k2 reduction presented in Table \ref{table_w} in the case of combining all four constraints and slightly larger values for the other cases.

Our estimates of the systematic uncertainty for MLCS2k2 are larger than the values calculated in \citet{kessler} because they use the BAO $A$-parameter from \citet{2005ApJ...633..560E}, and the added constraints on $\Omega_M$ and $w$ derived from using the CMB $R$-parameter \citep{2008arXiv0803.0547K}. We 
reproduce their values for the systematic uncertainties on MLCS2k2 ($\Delta w \approx 0.1$), if we include these constraints in our analysis. However, in this paper, we restrict our analysis to intermediate-redshift probes and therefore do not include the CMB constraints, which results in larger uncertainties.

Our analysis of the MLCS2k2 systematic uncertainties discussed above does not include the large shift in $w$ discussed in \citet{kessler} when the rest-frame $U$-band template is removed in the light-curve fitting. As seen in Table 6 of \citet{kessler}, removing the rest--frame U--band results in a $-0.31$ shift in $w$, while we find a shift of $-0.43$ if we remove this data. This particular uncertainty would therefore give rise to a bimodal result either centered around $w\simeq-0.8$ (with U--band included) or $w=-1.2$ (without U-band), yet both consistent with $w=-1$ within the errors.

We do not add the uncertainty due to excluding the rest--frame U--band to our systematic errors because we believe it is incorrect to exclude this data from the light--curve fitting. Even though there is evidence for diversity in the UV spectra of SNe Ia (see Ellis et al. 2008; Foley et al. 2008),  the removal of the rest-frame U--band data from the SDSS--only analysis results in the light curve fitter using only two filters at $z<0.2$ to constrain the colors of the SNe. This provides significant freedom to the MLCS2k2 fitter and then the priors on the fitted parameters become important. We note that $w$ is only shifted by $-0.1$ when using the SALT light curve fitter \citep[see Table 8 in][]{kessler} on the SDSS--only sample with the rest-frame U--band excluded.  This is the only noticeable difference between these two light curve fitting methodologies when considering the SDSS--only sample; namely the error on $w$ when the rest-frame U--band is excluded. Finally,  \citet{kessler} also sees a clear jump in the SDSS Hubble diagram at $z\simeq0.2$ when the rest-frame U--band is excluded from the MLCS2k2 analysis (see their Section 10.1.3 and Fig. 30), 
indicating that a constant $w$ model is not a good fit in this case.

\section{Conclusions}
\label{conclusions}

We present an analysis of the luminosity distances of Type Ia Supernovae from 
the Sloan Digital Sky Survey-II (SDSS-II) Supernova Survey in conjunction with other intermediate redshift ($z<0.4$) cosmological measurements including redshift-space distortions from the 2dFGRS, the ISW effect, and the BAO distance scale from both 
the SDSS and 2dFGRS. We have analyzed the SDSS-II SN luminosity distances using several 'model-independent' methods, including fitting the 
data using a $q(z)$ parameterization, principal components, and a non-parametric ``sliding window" method. 
We find consistent results between all these methods that provides evidence for an accelerating universe 
based solely on the first-year SDSS-II SN data. The strongest evidence we find comes when we make the strongest assumptions,
that $q_0$ is constant and the universe is flat which gives probability for acceleration of $>97\%$. 

We also compare our SDSS-II SN data with the local BAO measurements, and find they are in good agreement. This is in contrast with the findings of \citet{2007MNRAS.381.1053P} which found tension between the two 
distance measures, but confirms the new BAO analysis of Percival et al. (2009) who note that this tension has now lessened. Taking this observation further, we test the distance duality relation, i.e., for any metric theory of gravity, we expect 
$d_L/(d_A (1+z)^2) = 1$. We see no evidence for a discrepancy from this relation (at the one sigma level) in contrast to previous claims for a potential violation on the
$2\sigma$ level as seen in \citep{2004PhRvD..69j1305B, 2008JCAP...07..012L}.
Finally, we present a new measurement of the equation-of-state parameter of dark energy using a combination of geometrical distances in the universe and estimates for the growth rate of structure.
Our strongest constraint comes from the combination of all four data-sets discussed herein (SDSS-II SN, BAO, redshift-space distortions, ISW) with $w=-0.81^{+0.16}_{-0.18}(stat)$ and $\Omega_M=0.22^{+0.09}_{-0.08}(stat)$ (assuming a flat universe).  However, the combination of just the SDSS-II SNe and the ISW measurements alone is almost as powerful in constraining these parameters (Table 1). Our results only change slightly if we allow curvature to vary, consistent 
with the CMB measurements (see Appendix \ref{appendix_curvature}). We quote a systematic uncertainty of $\Delta w =\pm0.15$ based on the details of the MLCS2k2 light--curve fitter (see \citealt{kessler} for a fuller discussion). 

Thus we have shown that low-redshift cosmological probes give a self-consistent picture of the distance-redshift relation.  When combined with growth of structure and ISW at the same epoch that picture is consistent with $\Lambda$CDM and re-enforces the 
complementarity amongst other data and analyses in the literature.

\section*{Acknowledgements}

We thank an anonymous referee for helpful comments on this paper which greatly improved the content of the paper. 
RCN on behalf of the authors would like to thank Mike Turner for helpful discussions on the $q_0$ fit and
Eric Aubourg for useful discussions on the distance duality relation. We also thank Rick Kessler and Mark Sullivan for extensive discussions about their work and papers. TG thanks Jussi V\"{a}liviita for helpful suggestions. HL, CS and RCN are grateful to STFC for funding this research with rolling grants ST/F002335/1  and ST/H002774/1. 
H.-J.~S is supported by the D.O.E at Fermilab. A.V.F. is grateful for the support of US NSF grant AST--0607485.

Funding for the creation and distribution of the SDSS and SDSS-II
has been provided by the Alfred P. Sloan Foundation,
the Participating Institutions,
the National Science Foundation,
the U.S. Department of Energy,
the National Aeronautics and Space Administration,
the Japanese Monbukagakusho,
the Max Planck Society, and the Higher Education Funding Council for England.
The SDSS Web site \hbox{is {\tt http://www.sdss.org/}.}

The SDSS is managed by the Astrophysical Research Consortium
for the Participating Institutions.  The Participating Institutions are
the American Museum of Natural History,
Astrophysical Institute Potsdam,
University of Basel,
Cambridge University,
Case Western Reserve University,
University of Chicago,
Drexel University,
Fermilab,
the Institute for Advanced Study,
the Japan Participation Group,
Johns Hopkins University,
the Joint Institute for Nuclear Astrophysics,
the Kavli Institute for Particle Astrophysics and Cosmology,
the Korean Scientist Group,
the Chinese Academy of Sciences (LAMOST),
Los Alamos National Laboratory,
the Max-Planck-Institute for Astronomy (MPA),
the Max-Planck-Institute for Astrophysics (MPiA), 
New Mexico State University, 
Ohio State University,
University of Pittsburgh,
University of Portsmouth,
Princeton University,
the United States Naval Observatory,
and the University of Washington.

This work is based in part on observations made at the 
following telescopes.
The Hobby-Eberly Telescope (HET) is a joint project of the University of Texas
at Austin,
the Pennsylvania State University,  Stanford University,
Ludwig-Maximillians-Universit\"at M\"unchen, and Georg-August-Universit\"at
G\"ottingen.  The HET is named in honor of its principal benefactors,
William P. Hobby and Robert E. Eberly.  The Marcario Low-Resolution
Spectrograph is named for Mike Marcario of High Lonesome Optics, who
fabricated several optical elements 
for the instrument but died before its completion;
it is a joint project of the Hobby-Eberly Telescope partnership and the
Instituto de Astronom\'{\i}a de la Universidad Nacional Aut\'onoma de M\'exico.
The Apache 
Point Observatory 3.5 m telescope is owned and operated by 
the Astrophysical Research Consortium. We thank the observatory 
director, Suzanne Hawley, and site manager, Bruce Gillespie, for 
their support of this project.
The Subaru Telescope is operated by the National 
Astronomical Observatory of Japan. The William Herschel 
Telescope is operated by the 
Isaac Newton Group on the island of La Palma
in the Spanish Observatorio del Roque 
de los Muchachos of the Instituto de Astrofisica de 
Canarias. The W. M. Keck Observatory is operated as a scientific partnership 
among the California Institute of Technology, the University of 
California, and the National Aeronautics and Space Administration; the 
observatory was made possible by the generous financial support of the 
W. M. Keck Foundation. 

\appendix

\section{Cosmic Curvature}
\label{appendix_curvature}
\begin{figure}
\includegraphics[width=75mm] {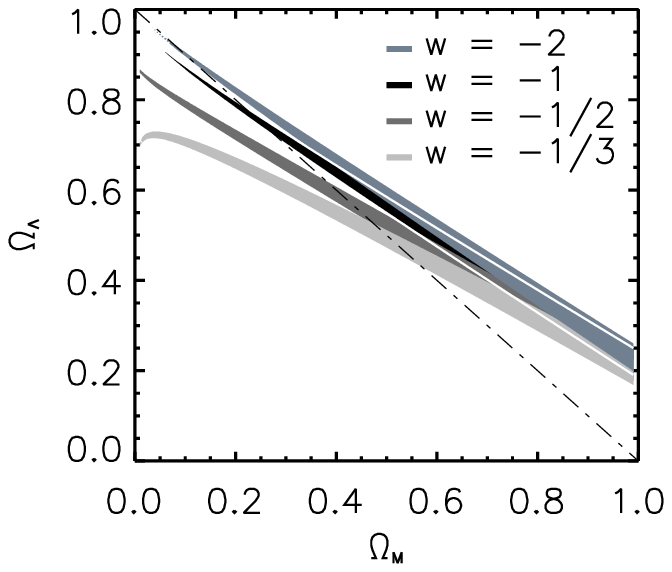}
 \caption{The $68\%$ contours in the $\Omega_{\Lambda}-\Omega_M$ plane derived from the ``scaled distance to recombination" $R$ (see text) taken from the WMAP-5 results, or $R=1.715 \pm 0.021$. The dot-dashed line indicates a flat universe.}
 \label{fig_curvature}
\end{figure}
In most parts of this paper, we assume a flat universe, or $\Omega_M + \Omega_{\Lambda} = 1$. Observationally,
the most robust constraints on curvature come from the distance to the last scattering surface of the CMB as determined by measurements of the CMB power spectrum. In cases where we consider curvature, allowing any combination of $\Omega_M$ and $\Omega_{\Lambda}$, we include a prior on the ``scaled distance to recombination" $R$ defined as
\begin{equation}
R=\sqrt{\Omega_M H_0^2} \frac{d_L(z_{\rm CMB})}{(1+z_{\rm CMB})},
\label{eq_r}
\end{equation}
given in \citet{2007PhRvD..76j3533W}, with a value of $R=1.71\pm 0.019$ from
\citet{2008arXiv0803.0547K}. We note that $R$ is independent of $H_0$ because $d_L$ scales linearly with 
$1/H_0$. For fixed values of $\Omega_M$ and $w$ it is now possible to constrain $\Omega_{\Lambda}$ from a measurement of $R$. 
We show $68\%$ confidence levels in Fig. \ref{fig_curvature} calculated from Eq. (\ref{eq_r}). As $R$ does not depend on the Hubble 
constant $H_0$, or on the baryon density $\Omega_b h^2$, no further assumptions are needed on these quantities. 
From Fig. \ref{fig_curvature}, it is obvious that using only $R$, a curvature of order a few percent cannot be neglected 
(depending on the value of $w$). We introduce the effects of curvature in the redshift-space distortion analysis by adding 
an additional $\chi^2$ term in the likelihood analysis calculated from Eq. (\ref{eq_r}) and subsequently marginalize 
over $\Omega_\Lambda$, where as for the ISW effect we pick for a given ($w, \Omega_M$) combination the best-fit 
value of $\Omega_\Lambda$.

\section{Correlated Peculiar Velocities}
\label{PV}

The peculiar velocities (PVs) of supernovae (SNe) introduce an additional scatter onto the Hubble diagram (see Eq. (\ref{eq_delta_z})). 
However, as pointed out by \cite{2006PhRvD..73l3526H}, we expect these PVs to be correlated, especially at low redshift, thus leading to 
significant covariance between pairs of SNe, i.e., a pair of SNe at radial positions $\mathbf{r}_i, \mathbf{r}_j$ have a projected velocity correlation 
function of $\xi(\mathbf{r}_i, \mathbf{r}_j) = \langle (\mathbf {v} (\mathbf{r}_i) \cdot \mathbf{\hat r}_i)  (\mathbf {v} (\mathbf{r}_j) \cdot \mathbf{\hat r}_j) \rangle$. 
We can calculate this function in linear theory using the matter power spectrum, the linear growth function and its derivative. This is interesting because, 
if this effect is detected, it may enable the SNe to constrain the parameters of structure formation, in addition to the standard background expansion.

The expression for the full covariance between SNe is given by \citet{2007PhRvL..99h1301G, 2008MNRAS.389L..47A}

\begin{equation} 
C_v(i,j) = \left(1 - \frac {(1+z)^2}{Hd_L}  \right)_i \left(1 - \frac {(1+z)^2}{Hd_L}  \right)_j  \xi (\mathbf r_i,\mathbf r_j).
\label{eq:cvm}
\end{equation}

This can be compared with the standard diagonal random errors, which are

\begin{equation} 
\sigma(i)^2 = \left( \frac{\ln 10}{5}  \right)^2 [\sigma_m^2 + \mu_{err}(i)^2] + \left(1 - \frac {(1+z)^2}{Hd_L}\right)_i \sigma_v^2,
\end{equation} 
where the intrinsic magnitude and velocity scatters $\sigma_m, \sigma_v$ have been introduced as usual.
A numerical evaluation shows that the two are comparable at low redshift. In particular, for a pair of SNe at $z = 0.05$ and zero angular separation, the covariance is $C_v(i,j) \simeq 0.1 \sigma(i) \sigma(j)$. This  decreases at higher redshifts and greater separations.

This effect has been detected by \cite{2007PhRvL..99h1301G} using a catalogue of 124 low redshift SNe by \cite{2007ApJ...659..122J} at $\bar z = 0.017$, and it has been carried further to constrain parameters such as $\sigma_8$ and the growth factor $\gamma$ \citep{2008MNRAS.389L..47A}.

Here we repeated the analysis for the SDSS SNe, but since our minimun redshift is $z\simeq 0.05 $, we expect the effect of correlated PVs to be small. Indeed, we found that a likelihood study performed by a Monte Carlo Markov chain (MCMC) analysis of the cosmological parameters yields no change in the results whether the full PV covariance matrix of Eq. (\ref{eq:cvm}) is included or not. For example, if we set  the intrinsic scatters $\sigma_m = \sigma_v = 0$, we find that the reduced $\chi^2 / \nu $ for the best fit cosmology decreases by only 1\% when the PV covariance matrix was included.

Therefore, we are unable to detect the correlation of SN peculiar velocities with this data and are safe to ignore them in further analyses. This effect will become more important with larger samples of low redshift SNe.  

\end{document}